\documentclass[letter]{aa}
\usepackage{graphicx}
\usepackage{txfonts}
\usepackage{svg}
\usepackage{mathrsfs}
\usepackage[percent]{overpic}
\usepackage{etoolbox}
\usepackage{xcolor} 

\usepackage[colorlinks=true,linkcolor=blue,citecolor=blue,filecolor=blue,urlcolor=blue]{hyperref}
\newcommand{\orcid}[1]{\unskip\protect\href{https://orcid.org/#1}{\protect\includegraphics[width=8pt,clip]{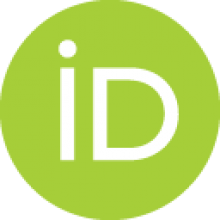}}}

\newcommand{\muHz}{\mu\mathrm{Hz}}
\def\mcC{{\mathcal{C}}}
\def\mcL{{\mathcal{L}}}
\def\mbbE{{\mathbb{E}}}
\def\mcP{{\mathcal{P}}}
\def\rmd{{\mathrm{d}}}

\newtoggle{showchanges}
\togglefalse{showchanges}   

\newcommand{\changes}[1]{%
  \iftoggle{showchanges}{\textbf{#1}}{#1}%
}
\defcitealias{Regulo-2016-A&A}{R16}
\defcitealias{Santos-2018-ApJS}{S18}
\begin{document}

\title{Determining low-$\ell$ p-mode frequency shifts in Sun-like stars}
\subtitle{Enhancing the cross-correlation technique with filters}
\titlerunning{cross-correlation}
\authorrunning{Kashyap et. al.}
\titlerunning{Filtered cross-correlation technique for asteroseismic p-modes}
\author{Samarth G. Kashyap\inst{1}\orcid{0000-0001-5443-5729} \and   Laurent Gizon\inst{1,2,3}\orcid{0000-0001-7696-8665} \and Jesper Schou\inst{1}\orcid{0000-0002-3997-2056}
\and Rachel Howe\inst{4}\orcid{0000-0002-3834-8585}}
\institute{Max-Planck-Institut f\"ur Sonnensystemforschung, Justus-von-Liebig-Weg 3, D-37077 G\"ottingen, Germany\\
\email{kashyap@mps.mpg.de} 
\and Institut f\"{u}r Astrophysik und Geophysik, Georg-August-Universität Göttingen, D-37077 G\"ottingen, Germany
\and Center for Space Science, NYUAD Institute, New York University Abu Dhabi, Abu Dhabi, UAE
\and School of Physics and Astronomy, University of Birmingham, Edgbaston, Birmingham B15 2TT, UK}
\date{Accepted for publication (A\&A Letters) on 25 Aug, 2026}
\abstract
{Acoustic mode frequencies in the Sun and Sun-like stars change due to magnetic activity on timescales that are much longer than that of the star's rotation and much shorter than that of its evolution. Given the poor S/N of the observed stellar p-modes, it is challenging to measure the changes of individual mode frequencies. Typically, power spectra of different time series segments are cross-correlated to estimate a mean p-mode frequency change, which ends up being an average of the individual mode contributions.}
{We seek to enhance the cross-correlation method by introducing a novel and computationally cheap method that enables us to disentangle p-mode frequency changes for different spherical harmonic degrees, $\ell$.}
{We designed filters that, if the inclination angle and rotation rate are already measured, enable the isolation of $\delta\omega_\ell$ and frequency changes of modes with a given $\ell$   while preventing bias creeping in from neighbouring modes. We performed Monte Carlo simulations to quantify the uncertainty in the estimation of $\delta\omega_\ell$.}
{We validated our method against well-studied solar data (from SOHO/VIRGO and BiSON) and demonstrate its applicability to the solar-like \textit{Kepler} star KIC 8006161.}
{}
\keywords{stars: interiors --- stars: magnetic field --- stars: oscillations (including pulsations) --- stars: solar-type}
\maketitle
\nolinenumbers
\section{Introduction}
Understanding stellar activity cycles is essential for studying the magnetic behavior of Sun-like stars and its implications for stellar evolution and planetary environments. The solar magnetic activity cycle can be observed from the frequency shifts of its pressure modes \citep[p-modes; e.g.,][]{Woodard-1985-Natur,AngueraGubau-1992-AnA}. Observations of other stars suggest that such cycles are common, with long-term spectroscopic and photometric studies detecting periodic variations in stellar activity \citep[for a review, see][]{Jeffers-2023-SSRv}. High-precision space-based photometry from missions such as \textit{CoRoT} (Convection, Rotation and Planetary Transits) and \textit{Kepler} has provided further evidence of activity cycles in thousands of stars \citep{Reinhold-2017-AnA}.

Helioseismology and asteroseismology studies have \changes{contributed to} this field by examining how oscillation frequencies, amplitudes, and linewidths change in response to magnetic activity \citep[e.g.,][]{Palle-1989-AAP,Chaplin-2003-ApJ,Salabert-2004-AAP}. The first confirmed detection of an activity cycle through asteroseismology was for the CoRoT star HD49933, where variations in mode parameters were observed over a few months \citep{Garcia-2010-Science,Salabert-2011-AAP}. Subsequently, the approach was extended to study other Sun-like stars observed by \textit{Kepler}, and the results of such studies reinforced the link between activity cycles and oscillation frequency shifts \citep{Salabert-2018-AAP,Santos-2018-ApJS}.
A key technique for detecting frequency shifts associated with stellar activity is cross-correlation (CC), where the ``average'' frequency shift of p-modes is determined by cross-correlating power spectra from two different time intervals. \changes{It was} developed for solar observations using single-site data \citep{Palle-1989-AAP} \changes{and} adapted for the study of Sun-like stars using space-based photometry \citep{Garcia-2010-Science,Mathur-2013-A&A}. \changes{\cite{Salabert-2011-AAP} attempted to separate odd and even degree modes with the CC method, but the resolution was limited by the proximity of these modes}. A significant advantage of this approach is that it does not require the precise extraction of individual mode frequencies, which is often difficult due to low S/Ns in short observational periods. Instead, by computing an overall shift in the power spectrum, the method enables robust tracking of activity-related variations \citep{Regulo-2016-A&A,Santos-2018-ApJS}. \changes{\cite{Kiefer-2017-AAP} highlighted KIC 8006161 as a solar analog with a strong activity cycle}. %

While concerns about the estimation of uncertainties have been addressed using Monte Carlo simulations \citep{Regulo-2016-A&A}, this method measures the ``average" frequency shift of all observed p-modes. To understand the latitudinal variation in activity, it is important to measure frequency shifts of individual modes, but the CC technique hides this information. As a first step, we seek a method that would work in the low-S/N regime without discarding information on the $\ell$ dependence of the mode frequency variation. 
In this work, we propose a novel extension to the CC technique: \changes{star-specific} filters, derived from fits to long-duration averaged spectra, isolate $\ell$-dependent frequency shifts in short, low-S/N segments without fitting every subseries. \changes{This method can be seen as a stepping stone for the quick inference of mode-by-mode frequency shifts in the low-S/N regime.}
\section{Method}
\begin{figure}
    \centering
    \includegraphics*[width=0.5\textwidth]{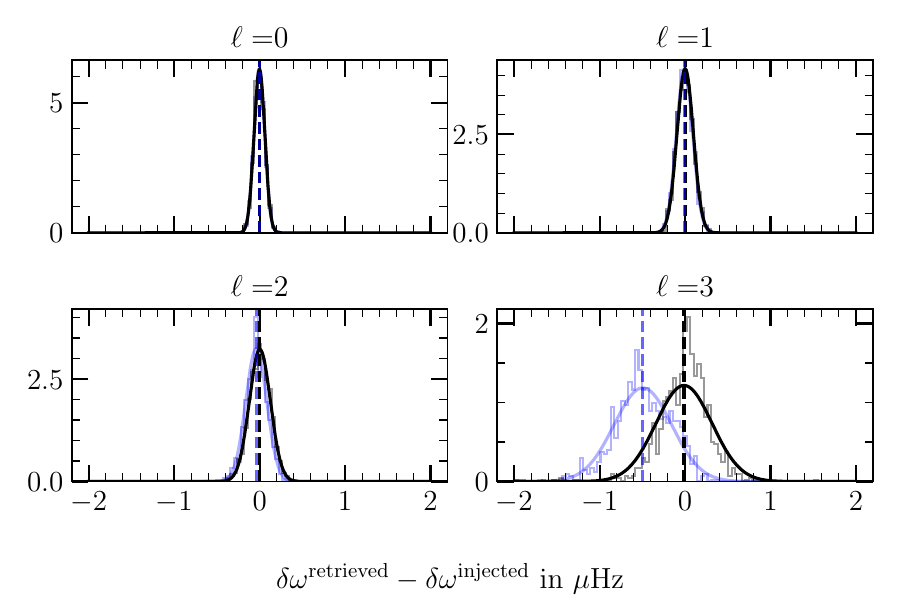}
    \caption{Monte Carlo simulation of the injection retrieval test. The blue (black) histograms correspond to results when the LC term is excluded (included).}
    \label{fig:two-filters-montecarlo}
\end{figure}

\begin{figure}
    \centering
    \includegraphics[width=1.0\linewidth]{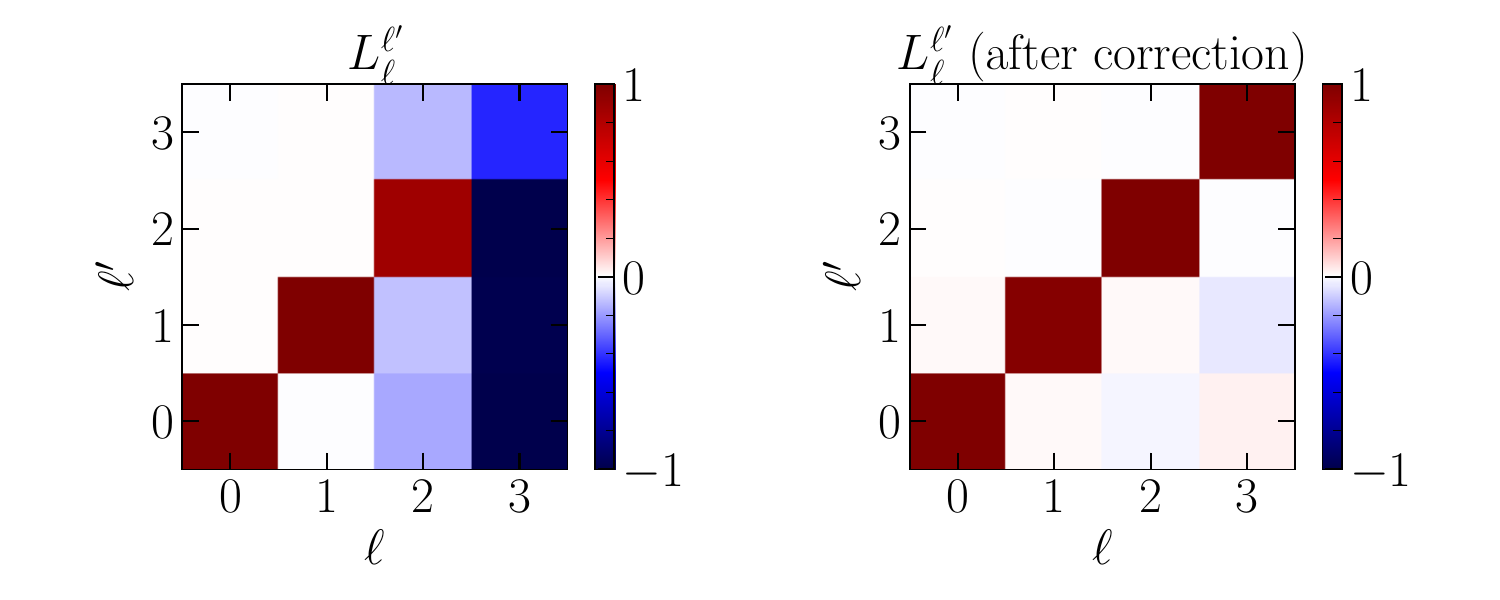}
    \caption{Leakage matrices computed before and after the introduction of the LC term.}
    \label{fig:leakage-matrix}
\end{figure}

In order to study activity cycles, the CC is typically performed with a reference power spectrum. \changes{While some works use the first subseries as the reference, we used the mean of all subseries to obtain the cleanest possible spectrum for constructing filters \citep[akin to][]{Regulo-2016-A&A}.} The procedure for constructing filters involves the following steps: 
\begin{enumerate}
    \item The observed time series is divided into subseries, $I_j(t)$. The corresponding power spectra, $P_j(\omega),$ are averaged to obtain the reference power spectrum, $P^\mathrm{ref}(\omega) = \frac{1}{j_\mathrm{max}}\sum_{j=1}^{j_\mathrm{max}} P_j(\omega)$.
    \item Peak-bagging is performed on the reference power spectrum, $P^\mathrm{ref}(\omega),$ to determine mode amplitudes, frequencies, and linewidths $(A_{n\ell}, \omega_{n\ell m}$, and $\Gamma_{n\ell}$) using a previously measured value of the inclination angle.
    \item Filters are constructed using the derived mode parameters.
    \item The power spectra, $P_j(\omega)$, are cross-correlated with the filters to estimate the frequency shift for a particular $\ell$.
\end{enumerate}
\changes{The CC method is computationally cheap and remains stable for short, low-S/N segments where full peak-bagging may fail.} The standard peak-bagging procedure involves fitting a model,
\begin{equation}
    P^\mathrm{ref}(\omega) = B(\omega) + \sum_{n\ell m} A_{n\ell m} \mcL(\omega;\omega_{n\ell m}, \Gamma_{n\ell}); \; A_{n\ell m} = A_{n\ell} V_{\ell m}(i),
    \label{eqn:peak-bag}
\end{equation}
where $B(\omega)$ is the non-seismic background and comprises two Harvey profile-like components \citep{Harvey-1985}, $A_{n\ell}$ is the mode amplitude, $V_{\ell m}(i)$ models the mode visibility changes due to the inclination angle, $i$ \citep{Gizon-Solanki-2003-ApJ}, and $\mcL(\omega; \omega_{n\ell m}, \Gamma_{n\ell})$ is a Lorentzian centered at the mode frequency $\omega_{n\ell m}$ and has a linewidth ($\Gamma_{n\ell}$) given by 
\begin{equation}
    \mcL(\omega; \omega_{n\ell m}, \Gamma_{n\ell}) = 
    \left[1 + [(\omega - \omega_{n\ell m})/(\Gamma_{n\ell}/2)]^2\right]^{-1}.
    \label{eqn:lorentzian}
\end{equation}
\changes{In this work, $\omega$ denotes the mode frequency (not the angular frequency).} The $m$ dependence of mode frequencies is modeled to come from an independent measurement of the mean rotation rate, $\Omega$:\begin{equation}
    \omega_{n\ell m} = \omega_{n\ell} + m\Omega (1 - C_{n\ell}) + \mathcal{O}(\Omega^2),
    \label{eqn:rotation-splittings}
\end{equation}
where $C_{n\ell}$ are the Ledoux constants \citep{Ledoux-1951-ApJ}, which account for the Coriolis force. The term that accounts for centrifugal distortion scales as $\Omega^2$ and is small when the rotation rate is comparable to that of the Sun and is thus ignored in the present work. While the solar inclination angle varies by $\pm 7.23^\circ$, we set $i=0^\circ$ for all subsequent calculations. The "mode-isolation" (MI) filter, denoted by $M^\ell(\omega),$ which is used to isolate modes corresponding to a specific $\ell$, is given by
\begin{equation}
    M^\ell(\omega) = B(\omega) + \sum_{n m} A_{n \ell m} \mcL(\omega; \omega_{n\ell m}, \Gamma_{n \ell}). \label{eqn:mi-filter}
\end{equation}
The MI filter is cross-correlated with the observed power spectrum to determine $\delta\omega_\ell$:
\begin{equation}
    \mcC^\ell_{j\mathrm{(MI)}}(\delta\omega) = \int P_j(\omega + \delta\omega) M^\ell(\omega) \, \rmd\omega.
    \label{eqn:mi-cross}
\end{equation}
While the MI filter targets modes of a specific $\ell$, the extraction is not perfect, owing to the proximity of neighboring modes. The bias in the measurement due to leakage from other $\ell$ channels is mitigated using the "leakage-correction" (LC) term, given by
\begin{align}
    \mcC^\ell_\mathrm{LC}(\delta\omega) & = \int M^\ell(\omega+\delta\omega) L^\ell(\omega) \, \rmd\omega, \label{eqn:lc-term} \\
    L^\ell(\omega) &= \sum_{n' \ell' m'; \ell'\neq \ell} A_{n' \ell' m'} \mcL(\omega; \omega_{n' \ell' m'}, \Gamma_{n' \ell'}). \label{eqn:lc-filter}
\end{align}
The LC term is small when modes are well separated and becomes significant when modes lie within a few linewidths of each other. The full expression of the filtered CC used to measure mean-frequency changes in a specific $\ell$ channel is\begin{equation}
    \mcC_j^\ell(\delta\omega) = \int P_j(\omega + \delta\omega) M^\ell(\omega) \, \rmd\omega - 
    \int M^\ell(\omega + \delta\omega) L^\ell(\omega) \, \rmd\omega.
    \label{eqn:cc-full}
\end{equation}
This CC function, $C^\ell(\delta\omega),$ is fitted with a Lorentzian, and its centroid is taken to be the estimate of $\delta\omega_\ell$. We performed an injection-retrieval test to establish the validity of the method: a frequency shift with a magnitude of 0.5 $\muHz$ was injected into all modes corresponding to a chosen $\ell$ and then retrieved using the proposed method. The $\delta\omega_\ell$ was estimated, first using only the MI filter and then introducing the LC term. Such a retrieval was performed for 1000 realizations of the spectrum. We note that using a higher number of simulated realizations does not change the final result. The results are summarized in Fig.~\ref{fig:two-filters-montecarlo}. We can see that the $\ell=3$ mode is the most affected by the correction, and therefore it is important to eliminate bias in the measurement of $\delta\omega$. The injection-retrieval test was generalized: $\delta\omega = 0.5$ $\muHz$ was injected into the $\ell$ channel and retrieved from the $\ell'$ channel. This enabled the quantification of cross-talk between different channels. We defined a leakage matrix, \(L_{\ell}^{\ell'} = \delta\omega_{\ell'}^\mathrm{retrieved}/\delta\omega_{\ell}^\mathrm{injected}\).
Figure~\ref{fig:leakage-matrix} shows the mean leakage matrix computed from the same 1000 realizations. It can be seen that, before the introduction of the LC term, there is significant leakage from the $\ell=2, 3$ channels into all the other channels. The LC term mitigates this and results in a diagonally dominant leakage matrix, with the off-diagonal terms smaller than their diagonal counterparts by two orders of magnitude.
\section{Method validation: Application to VIRGO data}
\label{sec:validation}
\begin{figure}
    \centering
    \includegraphics[width=0.9\linewidth]{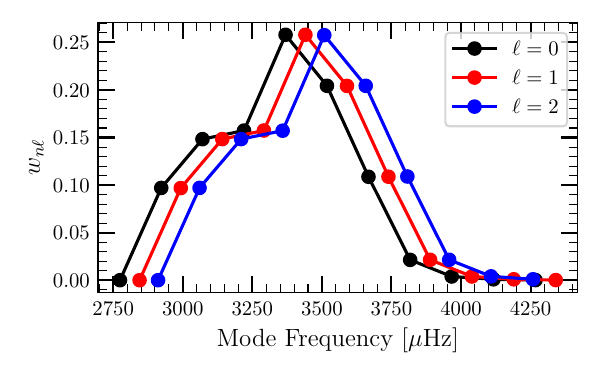}  
    \caption{Weights ($w_{n\ell}$; Eq.~\ref{eqn:weighted-delnu}) computed for KIC 8006161.}
    \label{fig:wnl-8006161}
\end{figure}

\begin{figure*}
    \centering
    \includegraphics[width=\linewidth]{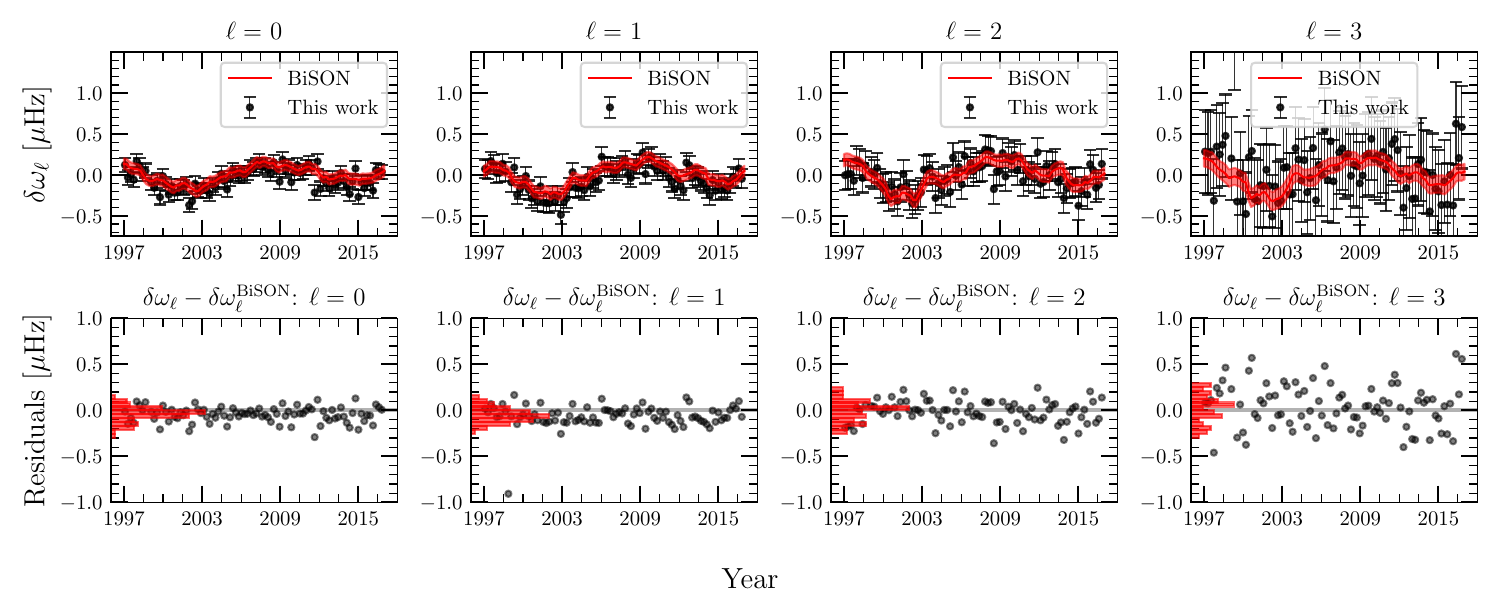}
    \caption{\textit{Top panels}: Peak-bagging BiSON spectra ($\delta\omega_\ell^\mathrm{BiSON}$; in red) with 1$\sigma$ errors  (shaded region), and the $\delta\omega_\ell$ estimated from the filtered CC technique (black dots with 1$\sigma$ error bars). \textit{Bottom panels}: Residuals (in black) and the corresponding normalized histogram (in red). The error bars for $\delta\omega_\ell$ were computed using Monte Carlo simulations.}
    \label{fig:delnu-ell-bison}
\end{figure*}

The analysis was applied to data from the VIRGO Sun photometers (SPM) onboard SOHO (Solar and Heliospheric Observatory) spanning solar cycles 23 and 24. VIRGO observes brightness variations of the Sun in the continuum. We analyzed the data from the red channel after de-trending and filling gaps with linear interpolation. To compare the results, mode frequencies from 108-day BiSON (Birmingham Solar Oscillations Network) time series segments were fitted using the method described in \cite{Howe-2023-MNRAS}, which yielded $\delta\omega_{n\ell}$ for modes with $\ell\le 3$. However, the CC technique provides us with an ``average'' frequency change. To enable comparison with a full spectral fitting, it is important to understand the quantity measured by the CC technique. In this section we show that the ``average'' frequency change $\delta\omega_\ell$ values have weights that depend on the amplitudes and linewidths of the observed modes. \changes{A derivation of the weights is provided in Appendix~\ref{apdx:weights}}. The weights ($w_{n\ell}$) are given by
\begin{equation}
    \delta\omega_\ell = \sum_n w_{n\ell} \; \delta\omega_{n\ell};\qquad 
    w_{n\ell} = A^2_{n\ell}/\Gamma_{n\ell} \left[\sum_{n} A^2_{n\ell}/\Gamma_{n\ell}\right]^{-1}.
    \label{eqn:weighted-delnu}
\end{equation}
We note that Eq. \ref{eqn:weighted-delnu}  is an approximation, as the fine structure of the spectrum has been ignored. Figure~\ref{fig:wnl-8006161} shows the weights computed for different modes. Given that the $w_{n\ell}$ scales as $A^2_{n\ell}/\Gamma_{n\ell}$, we see that $w_{n\ell}$ has a maxima around $\nu_\mathrm{max}$, the frequency corresponding to maximum power. The drop in $w_{n\ell}$ is less drastic for lower frequencies as both the amplitudes and linewidths are reduced. The peak-bagging results from BiSON data were weighted according to Eq.~\ref{eqn:weighted-delnu} and compared with the results derived using our method. Figure~\ref{fig:delnu-ell-bison} shows this comparison for the $\ell \le 3$ modes. While the CC technique captures the variation in frequencies, the residuals are not centered exactly at 0. This can be attributed to a mismatch in the reference frequencies. The reference for the CC method depends on the fits to the averaged spectra, and the reference for $\delta\omega_\ell^\mathrm{BiSON}$ has been taken to be the mean value of the observed time series in the chosen $\ell$ channel. The resulting bias is found to be on the order of 0.01 $\muHz$. The errors are seen to progressively grow for higher $\ell$. This provides a validation of our method for computing $\delta\omega_\ell$.
\section{Application to Sun-like stars}
\begin{figure}
    \centering
    \includegraphics[width=0.8\linewidth]{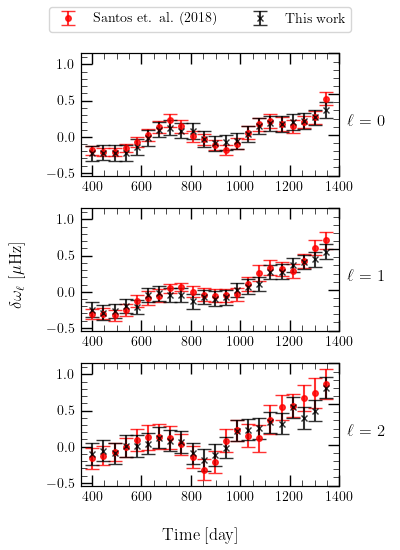}    
    \caption{Observed $\delta\omega_\ell$ of KIC 8006161 as reported by \cite{Santos-2018-ApJS} (in red) versus the results of the current method (in black). The errors were computed using Monte Carlo simulations.}
    \label{fig:delnu-ell-8006161}
\end{figure}

We extended the analysis to the Sun-like star KIC 8006161, whose seismic properties are similar to those of the Sun but has a significantly stronger magnetic activity cycle \citep[e.g.,][]{Kiefer-2017-AAP}. The light curve was obtained from the KASOC (Kepler Asteroseismic Science Operations Center)concatenated time series database\footnote{\href{https://kasoc.phys.au.dk/download.php?fileid=10989679&format=fits}{https://kasoc.phys.au.dk/download.php?fileid=10989679\&format=fits}}. Peak-bagging results derived via Monte Carlo techniques were taken  from \cite{Santos-2018-ApJS} (hereafter \citetalias{Santos-2018-ApJS}) and provide another source of validation for the current method. To construct the MI filter and LC term, the time series was divided into segments of 90 days in length, with two neighboring segments having an overlap of 45 days. Peak-bagging was first performed on $P^\mathrm{ref}(\omega)$ using the \texttt{apollinaire} package (\citealt{Breton-2022-AAP}; see our Appendix~\ref{sec:apdx:peakbag-8006161}). An inclination angle ($i$) of $40^\circ$ and a rotation rate ($\Omega$) of $0.54\muHz$ \citep{Kamiaka-2018-MNRAS} were used to construct the MI filters and the LC terms. Figure~\ref{fig:delnu-ell-8006161} shows the comparison of $\delta\omega_\ell$ estimates from the current method with those from the peak-bagging analysis of \citetalias{Santos-2018-ApJS}. It is seen that the frequency shifts estimated using the proposed technique broadly agree with the  \citetalias{Santos-2018-ApJS} results, barring a few outliers at later times.
\section{Discussion and conclusion}
The CC technique, while computationally cheap, hides information regarding the dependence of $\ell$  on frequency. We extended the method by constructing filters that can isolate mode frequency variations of specific $\ell$. We also introduced a correction term to overcome bias in the filter. A simple analytical derivation helps map individual mode frequency variations to the measurement from the filtered CC technique. Monte Carlo simulations help quantify the uncertainties in the measurement of $\delta\omega_\ell$.

The current estimation of the weights does not consider the $m$-splitting of frequencies. While a comparison with both VIRGO and \textit{Kepler} data suggests that the approximation is broadly valid, the outliers warrant a more elaborate derivation of weights that considers the amplitude variation of the multiplets too. Another limitation of the current method is the focus around the $\nu_\mathrm{max}$ region since it is known that higher-frequency modes  also have higher variabilities due to changes in activity. Designing better filters, where the weights for the high-frequency modes are not small, could improve the significance of the estimated $\delta\omega_\ell$. The current analysis also assumes that the inclination angle and the rotation rate of the star have already been measured independently. Seismic determinations of rotation rates as well as inclination angles need to be included within the pipeline for a successful application to a larger dataset, as well as for data from future missions.
Our analysis of KIC 8006161 shows that the method is applicable for \textit{Kepler} data as well. The length of the subseries is dependent on the length of the activity cycle, along with the total observation time. The estimation of errors is also validated as they broadly agree with errors computed from the peak-bagging of individual subseries. The CC method captures the variations in frequencies well; however, owing to differences in the way the references are constructed, the $\delta\omega_\ell$ for different methods are only consistent up to a constant. This in itself is not a concern as the process of constructing the reference has been specified and the same process can be used for any downstream inferences, such as active latitude estimation in stars.\\

\section*{Software availability} The \textsc{seismo-xl} code developed for this work is publicly available on GitHub at \url{https://github.com/samarth-kashyap/seismo-xl} and is archived on Zenodo \citep{kashyap_2026_22085865}. It is released under the MIT license.

\begin{acknowledgements}
L.G. acknowledges funding from project BUTTERFLY (DFG grant 530101854). R.H. acknowledges the support of the UK Science and Technology Facilities Council (STFC) through grant ST/V000500/1. The computations in this paper related to BiSON data were performed using the University of Birmingham’s BlueBEAR HPC service, which provides a High Performance Computing service to the University’s research community.
See \href{http://www.birmingham.ac.uk/bear}{http://www.birmingham.ac.uk/bear} for more details.
We would like to thank all those who have been associated with BiSON over the years. We also thank \^{A}ngela R. G. Santos for providing the peak-bagging results of KIC 8006161 from \citetalias{Santos-2018-ApJS}.
\end{acknowledgements}
\bibliography{biblio}{}
\bibliographystyle{aa}

\begin{appendix}
\section{MI filters}
\label{sec:apdx:mi-filters}
\begin{figure*}
    \centering
    \includegraphics[width=1.0\linewidth]{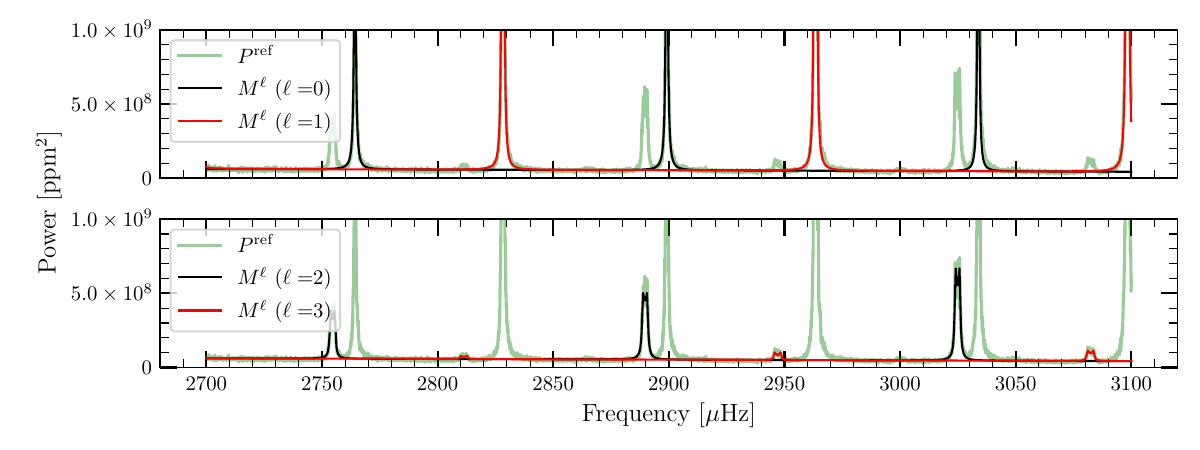}   
    \caption{Reference power spectrum, $P^\mathrm{ref}(\omega),$ computed from \changes{VIRGO} observations (in green). The MI filters for odd $\ell$ are shown in red and those for even $\ell$  in black. For the sake of clarity, only a small frequency range is shown.}
    \label{fig:mi-filters}
\end{figure*}

Figure~\ref{fig:mi-filters} shows the MI filters computed from VIRGO/SPM power spectrum. This spectrum has visibility of modes for $\ell \le 3$ and hence four different filters are constructed.
\section{Fitting the cross-correlation function}
\label{sec:apdx:ccfit}
\begin{figure*}
    \centering
    \includegraphics[width=0.9\linewidth]{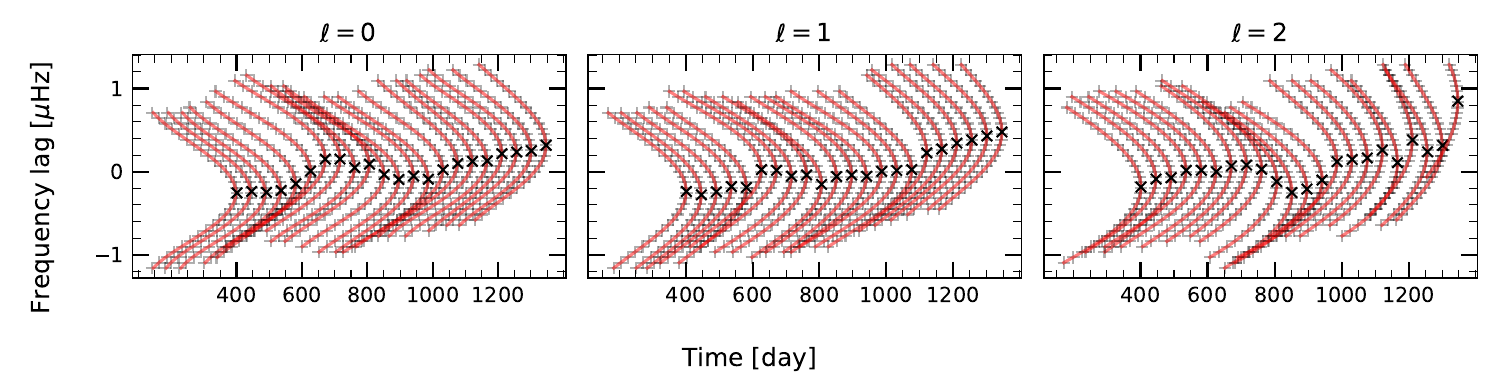}
    \caption{Fit for the CC function for KIC 8006161. In each panel, the black "+" show the $C^\ell(\delta\omega)$ computed using Eq.~\ref{eqn:cc-full} and the red curves show their best Lorentzian fits. The cross indicates the location of the peak. The $C^\ell(\delta\omega)$ have been scaled and shifted so that the locations of the peaks coincide with the observation time.}
    \label{fig:ccfit}
\end{figure*}

\begin{figure*}
    \centering
    \includegraphics[width=\linewidth]{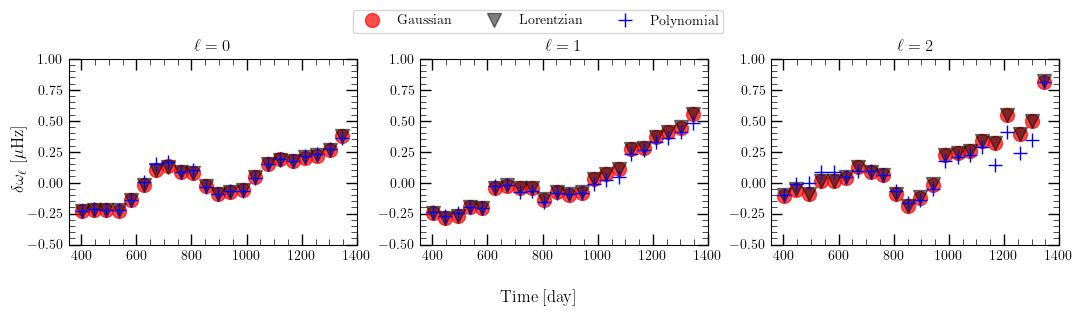}
    \caption{Comparison of frequency shifts measured by fitting three different functional forms.}
    \label{fig:comparison-methods}
\end{figure*}

Figure~\ref{fig:ccfit} shows the CC function and the corresponding best Lorentzian fit. 
It is seen that the fits are excellent and provide a reliable measure of $\delta\omega_\ell$. We have also tried fitting the curve with a Gaussian \citep{Regulo-2016-A&A} and with a quadratic (based on the approximation to Eq.~\ref{eqn:cc-expr}). The final result is found to be robust. \changes{Figure~\ref{fig:comparison-methods} shows the frequency changes measured by fitting the 3 functions above. It can be seen that the results are stable across different functions, with quadratic function being more prone to being affected by outliers, which is especially pronounced for $\ell=2$.}.
\section{Statistics of the cross-correlation function}
\label{apdx:sec:cc-statistics}
The CC method described in this work involves cross-correlating a noisy power-spectrum ($\chi^2_2-$statistics), with a filter $M^\ell(\omega)$ which is a smooth function. \changes{The CC function can be written as}
\begin{align}
    \mcC^\ell(\delta\omega) &= \sum_k \, P_j(\omega_k) M^\ell(\omega_k+\delta\omega) \chi^2_2(\omega_k) \nonumber \\ 
    &= \sum_k \mcP^\ell_j(\omega_k+\delta\omega) \chi^2_2(\omega_k),
\end{align}
where $P_j(\omega_k) \chi^2_2(\omega_k)$ is the noisy power spectrum and $\mcP^\ell_j(\omega_k) = P_j(\omega_k)M^\ell(\omega_k)$. From the above equation, it is evident that the CC function $\mcC^\ell$ is a linear combination of random variables distributed according to the $\chi^2_2$ distribution. The total number of frequency bins $(\omega_k)$, depends on the length of the subseries used in the analysis. For a subseries length of 30 days with an observation cadence of 1 minute, we have 43200 frequency bins. Since we are dealing with a summation of a large number of random variables, we consider the applicability of the central limit theorem (CLT). These random variables are independent but not identical distribution and hence we need to consider the applicability of the Lyapunov CLT. The Lyapunov CLT states that a summation of a large number of random variables results in a variable with a zero-mean Gaussian distribution, if the Lyapunov condition is satisfied. The Lyapunov condition for a sequence of independent random variables $X_k$ for $k=1, 2, ... n$ with expectation values given by $\mu_k = \mbbE[X_k]$, then for some $\delta>0$, we have
\begin{equation}
    \mathrm{Lyp} = \lim_{n\to \infty}  \frac{\sum_{k=1}^n\mbbE\left[|X_k - \mu_k|^{2+\delta}\right]}{\left(\sum_{k=1}^n \mbbE\left[|X_k - \mu_k|^2\right]\right)^{1+\delta/2}} = 0.
    \label{eqn:lmlt}
\end{equation}
Since the observed power-spectrum has a $\chi^2_2-$distribution, we have 
\begin{equation}
    \mbbE[\alpha \chi^2_2] = \mu = \alpha.
\end{equation}
Rewriting the Lyapunov condition for the specific case, we have 
\begin{equation}
    \mathrm{Lyp} = \lim_{n\to \infty}  
    \frac{\sum_{k=1}^n\mbbE\left[|\mcP^\ell_j(\omega_k) \chi^2_2 - \mcP^\ell_j(\omega_k)|^{2+\delta}\right]}{\left(\sum_{k=1}^n \mbbE\left[|\mcP^\ell_j(\omega_k) \chi^2_2 - \mcP^\ell_j(\omega_k)|^2\right]\right)^{1+\delta/2}}
\end{equation}
It can be seen that the expression scales as $n$. We can compute the above term for, say, $\delta=1$. The numerator and denominators of the Lyapunov condition can be shown to be 
\begin{align}
    \mathrm{Nr}_1 & =  2^4 \sum_{k=1}^n \mcP_j^\ell(\omega_k)^3 \nonumber \\ 
    \mathrm{Dr}_1 & = 2^2 \sum_{k=1}^n \mcP_j^\ell(\omega_k)^2
\end{align}
To understand the behaviour of the numerator and denominator in the limit of large $n$, we first consider the case when the star is observed for twice the duration. This results in a refinement of the frequency resolution i.e., doubling the observation time improves the frequency resolution by a factor of 2. These terms for the power spectrum can be written as 
\begin{align}
    \mathrm{Nr}_2 & =  2^4 \sum_{k=1}^{2n} \mcP_j^\ell(\omega_k)^3 \approx 2 \mathrm{Nr}_1 \nonumber \\ 
    \mathrm{Dr}_2 & = 2^2 \sum_{k=1}^{2n} \mcP_j^\ell(\omega_k)^2 \approx 2\mathrm{Dr}_1
\end{align}
Assuming that the spectra of such a star is unchanged, the above terms on the refined frequency grid can be approximated to be the nearest neighbour interpolation on the unrefined grid. This implies that, increasing the observation time by a factor of $\lambda$ results in 
\begin{align}
    \mathrm{Nr}_\lambda & =  \lambda \mathrm{Nr}_1 \nonumber \\ 
    \mathrm{Dr}_\lambda & = \lambda \mathrm{Dr}_1.
\end{align}
The Lyapunov condition can now be written as 
\begin{equation}
    \mathrm{Lyp} = \lim_{\lambda\to\infty} \frac{\lambda \mathrm{Nr}_1}{(\lambda\mathrm{Dr}_1)^{1.5}}.
\end{equation}
Since the power in the signal is finite, both $\mathrm{Nr}_1$ and $\mathrm{Dr}_1$ are finite. Hence $\mathrm{Lyp}$ scales as $\lambda^{-0.5}$ which vanishes for large $\lambda$. This implies that Lyapunov CLT is applicable for the CC function i.e., the statistics are Gaussian.
\section{Peak-bagging KIC 8006161}
\label{sec:apdx:peakbag-8006161}
\begin{figure*}
    \centering
    \includegraphics[width=1\linewidth]{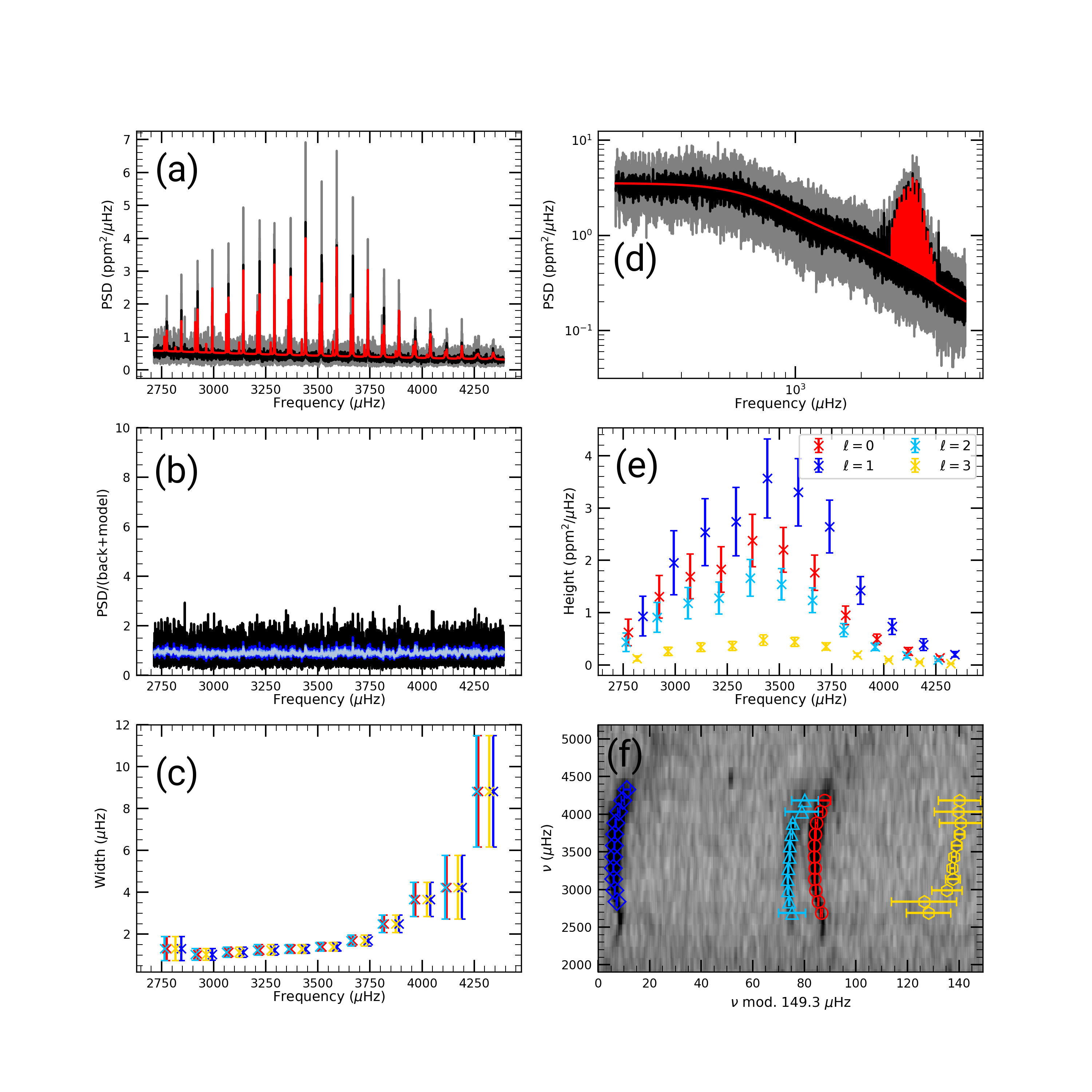}
    \caption{Peak-bagging summary of KIC 8006161. \textit{Panels (a) and (d)}: Observed $P^\mathrm{ref}(\omega)$ in gray, smoothed $P^\mathrm{ref}$ in black, and model $P$ in red. \textit{Panel (b)}: $P^\mathrm{ref}(\omega)/P^\mathrm{model}(\omega)$ for observed data in black and smoothed data in blue. \textit{Panels (e), (c), and (f)}: Frequencies, linewidths, and the \'echelle diagram.}
    \label{fig:peakbag-8006161}
\end{figure*}

Figure~\ref{fig:peakbag-8006161} shows the summary of peak-bagging process for KIC 8006161 using the \texttt{apollinaire} package. The frequencies computed here are used to construct the MI-filters and their corresponding LC terms. These frequencies are the reference mode frequencies about which variations are measured.
\section{Derivation of weights}
\label{apdx:weights}
\changes{The proposed CC method combines the frequency shifts of modes with different radial orders $n$ at a given spherical harmonic degree $\ell$. In this section we show that the measure frequency shift $\delta\omega_\ell$ is a weighted-average of the shifts $\delta\omega_{n\ell}$. To compute these weights, we consider a simple model power spectrum given by 
\begin{equation}
    P(\omega) = \sum_{n\ell} A_{n\ell} \mcL_{n\ell}(\omega; \omega_{n\ell}, \Gamma_{n\ell}).
\end{equation}
Assuming the modes to be well separated and retaining only the dominant contributors to the CC function, we have
\begin{align}
    \mcC^\ell(\delta\omega) = \sum_{n} A^2_{n\ell} & \int
    \Big[\mcL(\omega+\delta\omega; \omega_{n\ell}, \Gamma_{n\ell}) \times \nonumber\\
    & \mcL(\omega; \omega_{n\ell} +\delta\omega_{n\ell}, \Gamma_{n\ell})\Big] \, \rmd\omega.
\end{align}
Using the expression of Lorentzian from Eq.~\ref{eqn:lorentzian}, we have
\begin{align}
    \mcC^\ell(\delta\omega) = \sum_n \pi A_{n\ell}^2 \Gamma_{n\ell}/4 \left[1 + \left((\delta\omega - \delta\omega_{n\ell})/\Gamma_{n\ell}\right)^2 \right]^{-1}.
    \label{eqn:cc-expr}
\end{align}
In the neighbourhood of the maximum, $x = (\delta\omega - \delta\omega_{n\ell})/\Gamma_{n\ell}$ is small. For small $x$, $1/(1+x^2)$ can be approximated to be $(1-x^2)$. Using this approximation, it can be shown that the maximum of $C^\ell(\delta\omega)$ is located at $\delta\omega_\ell$ given by
\begin{equation}
    \delta\omega_\ell = \sum_n w_{n\ell} \; \delta\omega_{n\ell};\qquad 
    w_{n\ell} = A^2_{n\ell}/\Gamma_{n\ell} \left[\sum_{n} A^2_{n\ell}/\Gamma_{n\ell}\right]^{-1},
    \label{eqn:weighted-delnu-2}
\end{equation}
where $w_{n\ell}$ are the weights.}
\end{appendix}
\end{document}